%% file: mnras_template.tex
\documentclass[fleqn,usenatbib]{mnras}

\usepackage{newtxtext,newtxmath}
\usepackage{amsfonts}
\usepackage{amsmath}
\usepackage{graphicx}
\usepackage{subfig}
\usepackage{url}
\usepackage{hyperref}
\usepackage{graphicx}

\usepackage[T1]{fontenc}

\DeclareRobustCommand{\VAN}[3]{#2}
\let\VANthebibliography\thebibliography
\def\thebibliography{\DeclareRobustCommand{\VAN}[3]{##3}\VANthebibliography}

\usepackage{graphicx}	% Including figure files
\usepackage{amsmath}	% Advanced maths commands
\usepackage[table,xcdraw]{xcolor}
\title[Missed Galactic novae in WISE]{Rapidly evolving Galactic plane outbursts in NEOWISE: Revisiting the Galactic nova rate with the first all-sky search in the mid-infrared}

\author[Zuckerman et al.]{
Liam Zuckerman$^{1}$\thanks{E-mail: lizuckerman@packer.edu}, Kishalay De$^{2,3}$, Anna-Christina Eilers$^2$,  Aaron M. Meisner$^{4}$, and Christos Panagiotou$^2$\\
$^{1}$ Packer Collegiate Institute, 170 Joralemon St, Brooklyn, NY 11201, USA\\
$^{2}$MIT-Kavli Institute for Astrophysics and Space Research, 77 Massachusetts Ave., Cambridge, MA 02139, USA\\
$^{3}$NASA Einstein Fellow\\
$^{4}$NSF's National Optical-Infrared Astronomy Research Laboratory, 950 N. Cherry Ave., Tucson, AZ 85719, USA
}

\date{Accepted XXX. Received YYY; in original form ZZZ}

\pubyear{2015}

\begin{document}
\label{firstpage}
\pagerange{\pageref{firstpage}--\pageref{lastpage}}
\maketitle

% Abstract of the paper
\begin{abstract}
The Galactic nova rate is intimately linked to our understanding of its chemical enrichment and progenitor channels of Type Ia supernovae. Yet past estimates have varied by more than an order of magnitude, ($\approx10-300$\,yr$^{-1}$) owing to limitations in both discovery methods as well as assumptions regarding the Galactic dust distribution and extragalactic stellar populations. Recent estimates utilizing synoptic near-infrared surveys have begun to provide a glimpse of a consensus ($\approx25-50$\,yr$^{-1}$); however, a consistent estimate remains lacking. Here, we present the first all-sky search for Galactic novae using 8 years of data from the NEOWISE mid-infrared (MIR) survey. Operating at $3.4$ and $4.6$\,$\mu$m where interstellar extinction is negligible, the 6-month cadence NEOWISE dataset offers unique sensitivity to discover slowly evolving novae across the entire Galaxy. Using a novel image subtraction pipeline together with systematic selection criteria, we identify a sample of 49 rapidly evolving MIR outbursts as candidate Galactic novae. While 27 of these sources are known novae, the remaining are previously missed nova candidates discovered in this work. The unknown events are spatially clustered along the densest and most heavily obscured regions of the Galaxy where previous novae are severely underrepresented. We use simulations of the NEOWISE survey strategy, the pipeline detection efficiency, and our criteria to derive a Galactic nova rate of $47.9^{+3.1}_{-8.3}$\,yr$^{-1}$. The discovery of these exceptionally bright (yet overlooked) nova candidates confirm emerging suggestions that optical surveys have been highly incomplete in searches for Galactic novae, highlighting the potential for MIR searches in revealing the demographics of Galactic stellar outbursts.
\end{abstract}

% Select between one and six entries from the list of approved keywords.
% Don't make up new ones.

%%%%%%%%%%%%%%%%%%%%%%%%%%%%%%%%%%%%%%%%%%%%%%%%%%

%%%%%%%%%%%%%%%%% BODY OF PAPER %%%%%%%%%%%%%%%%%%

\section{Introduction}

A white dwarf (WD) accreting mass from a binary companion can undergo thermonuclear eruptions caused by unstable, nuclear burning in novae \citep{2008clno.book.....B,Starrfield_2016, DellaValle2020, Chomiuk_2021}. Not only are these eruptions important phases in the evolution of lower-mass binary stars, they also contribute substantially to the nucleosynthesis of isotopes like  $^7$Li, $^{22}$Na, $^{26}$Al and $^{15}$N \citep{Gehrz_1998, JOSE2006550, 10.1046/j.1365-8711.2003.06526.x, Prantzos2012}. While WDs can gain mass via long-term accretion from their companions, potentially reaching the Chandrasekhar mass as progenitors of Type Ia supernovae, mass loss during nova eruptions can significantly alter the fate of the WD \citep{Soraisam2015, Starrfield_2020, doi:10.1146/annurev-astro-082812-141031,DARNLEY20201147}. Together, the physics and demographics of novae in our Galactic backyard have long been considered to be a cornerstone of stellar evolution, cosmology, and cosmic nucleosynthesis.

Although novae are some of the brightest stellar eruptions in the local universe, the nova rate in the Milky Way has remained uncertain---even after decades of investigations. Previous estimates rely on two independent methods. The first, and more widely used technique, involves extrapolating nova rates estimated for other galaxies via their $K$-band luminosity to the Milky Way, which produces a range of results spanning $ \approx 10-50 \text{ yr}^{-1}$ \citep{10.1093/mnras/stac2960, 1990AJ.....99.1079C,van_den_Bergh_1991,Shafter_2000,darnley,Rector2022}. The second class of methods relies on statistics of Galactic novae discovered in optical searches, resulting in estimates ranging from $\approx 30 - 300 \text{ yr}^{-1}$ \citep{10.1093/mnras/114.4.387,Liller_1987,hatano,Shafter_1997,ozdonmez}. While the former method suffers from differences in the stellar populations (e.g. star formation history and Galactic structure) between the Milky Way and external galaxies, the latter method remains limited because of uncertainties on the severe dust extinction and probed optical volume along the Galactic plane. While \citet{Mr_z_2015} provide a rate estimate based on a controlled set of nova candidates from the OGLE survey, the sample is not complete with spectroscopic classifications. Similarly, the Vista Variables in the Via Lactea (VVV; \citealt{Catelan2011}) survey reported $\approx 20$ dust-obscured nova candidates (e.g. \citealt{Saito2012, Saito2013a, Saito2013b, ContrerasPena2017a, ContrerasPena2017b}) from a search for large amplitude outbursts; however, without spectroscopic confirmation, this sample is also contaminated with young stellar object (YSO) outbursts and foreground dwarf novae \citep{ContrerasPena2017a}.

The advent of all-sky time domain surveys in the optical and near-infrared (NIR) bands has begun to change the landscape of nova demographics. \citet{De_2021} carried out the first systematic and spectroscopically complete search for novae in the Galactic plane ($\delta > -30^\circ$) using the near-infrared Palomar Gattini-IR (PGIR; \citealt{De2020}) survey. By carrying out complete simulations of the survey scheduling and detection efficiency, they estimated a Galactic nova rate of $43.7^{+19.5}_{-8.7}$\,yr$^{-1}$. In particular, they showed that previous optical searches were systematically missing a substantial population of dust-obscured novae, resulting in inconsistencies between observed and estimated nova rates in optical searches. More recently, \citet{Kawash_2021} and \citet{Kawash_2022} used novae discovered in the ASASSN and Gaia surveys to carry out a systematic analysis of the nova rate, providing a rate of $26\pm 5$\,yr$^{-1}$. With the estimated optical nova rate nominally inconsistent with the higher NIR rate, further progress requires additional investigation of independent surveys and observing wavebands to assess these discrepancies.

In this paper, we use data acquired with NASA's Wide-field Infrared Survey Explorer (WISE) space telescope \citep{Wright2010} to carry out the first systematic search for large-amplitude and rapidly evolving outbursts near the Galactic plane in the mid-infrared bands. Because WISE has been running for more than a decade, surveying the whole sky with a $\approx 6$ month cadence, it constitutes a promising, uniform time--domain survey to revisit the demographics of Galactic novae. Operating in the mid-infrared bands makes it effectively immune to dust extinction that affects both optical and (to a lesser extent) previous NIR searches, while the long-time baseline and whole-sky coverage provides a novel search method that is completely distinct from previous methods. The paper is structured as follows. In Section \ref{sec:identification}, we detail the process and criteria for selecting the candidate novae. Then, we analyze the characteristics of nova candidates in Section \ref{sec:characteristics}. In Section \ref{sec:rate}, we outline the procedures for determining the Galactic nova rate implied by the WISE data, and we present our nova-rate estimate. In Section \ref{sec:discussion}, we discuss the implications of the population of missed nova candidates identified in WISE and the overall Galactic rate. We summarize our findings in Section \ref{sec:summary}.

\section{WISE Sample of Candidate Novae}
\label{sec:identification}
%description of what the pipeline does

The Wide-field Infrared Survey Explorer (WISE) satellite \citep{Wright2010}, re-initiated as the NEOWISE mission \citep{Mainzer2014}, has been carrying out an all-sky mid-infrared (mid-IR) survey in the $W1$ ($3.4$\,$\mu$m) and $W2$ ($4.6$\,$\mu$m) bands since 2014. As part of an ongoing program to identify large amplitude mid-IR transients in NEOWISE data, we carried out a systematic search for transients in time-resolved co-added images created as part of the unWISE project \citep{Lang2014, Meisner2018}. The details of this search will be presented in De et al. (in prep). In brief, we used a customized code \citep{De2020} based on the ZOGY algorithm \citep{Zackay2016} to perform image subtraction on the NEOWISE images using the full-depth co-added images of the WISE mission (obtained during 2010-2011) as reference images\footnote{For all transients identified in the WISE Transient Pipeline (WTP, De in prep.), we adopt the naming scheme WTP\,XXYYYYYY, where XX indicates the year of first detection and YYYYYY is a six letter alphabetical code.}. Our pipeline produces a database of all transient sources down to a statistical significance of $\approx 10\sigma$. 

\begin{figure*}
    \centering
    \includegraphics[width=0.245\textwidth]{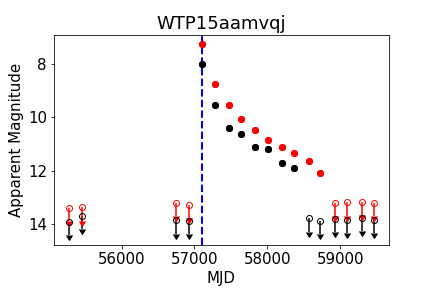}
    \includegraphics[width=0.245\textwidth]{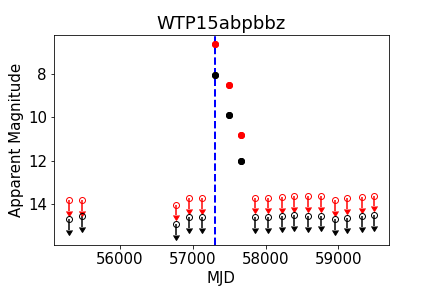}
    \includegraphics[width=0.245\textwidth]{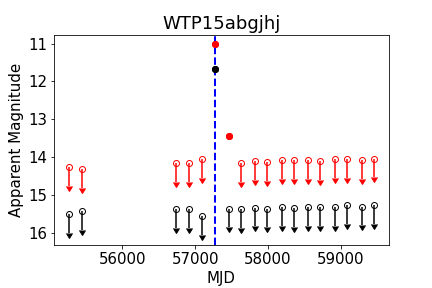}
    \includegraphics[width=0.245\textwidth]{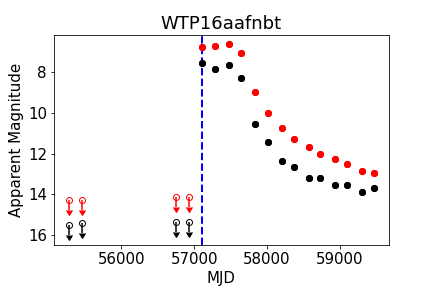}
    \includegraphics[width=0.245\textwidth]{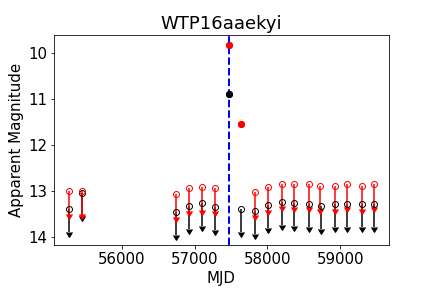}
    \includegraphics[width=0.245\textwidth]{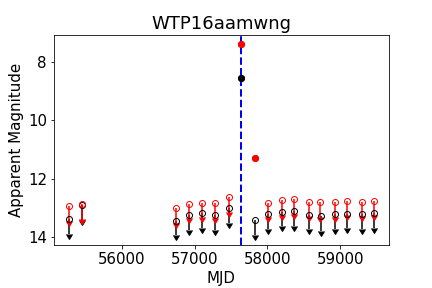}
    \includegraphics[width=0.245\textwidth]{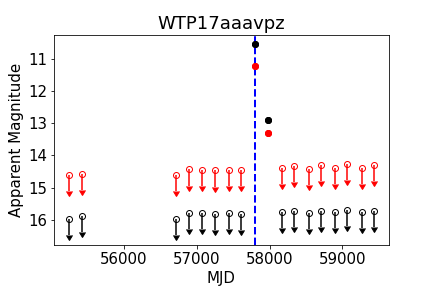}
    \includegraphics[width=0.245\textwidth]{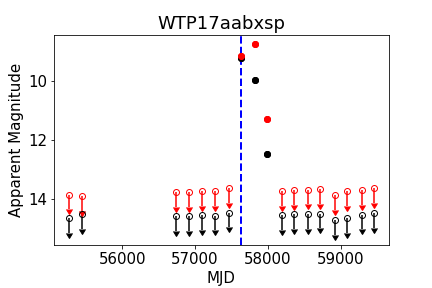}
    \includegraphics[width=0.245\textwidth]{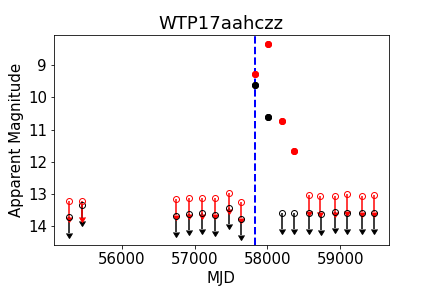}
    \includegraphics[width=0.245\textwidth]{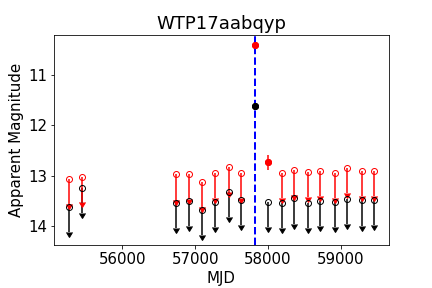}
    \includegraphics[width=0.245\textwidth]{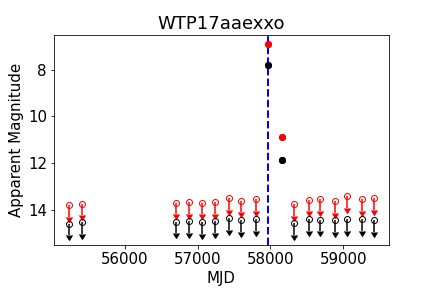}
    \includegraphics[width=0.245\textwidth]{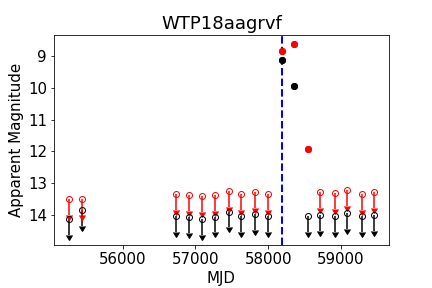}
    \includegraphics[width=0.245\textwidth]{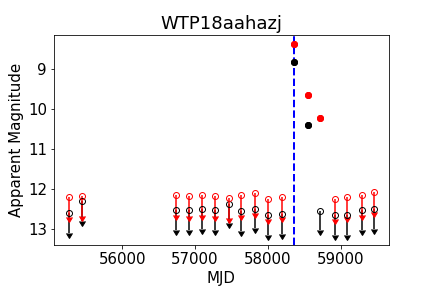}
    \includegraphics[width=0.245\textwidth]{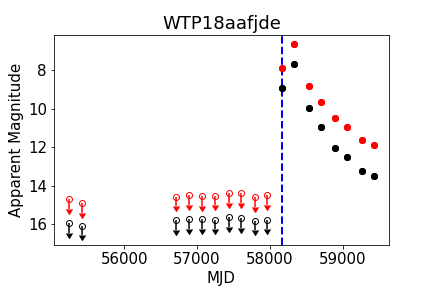}
    \includegraphics[width=0.245\textwidth]{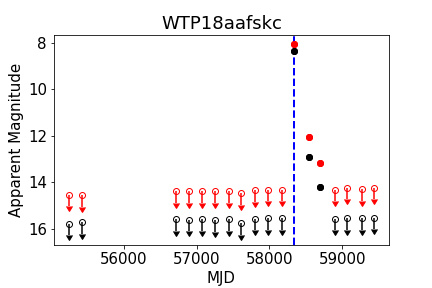}
    \includegraphics[width=0.245\textwidth]{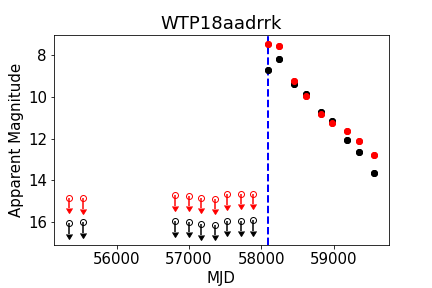}
    \includegraphics[width=0.245\textwidth]{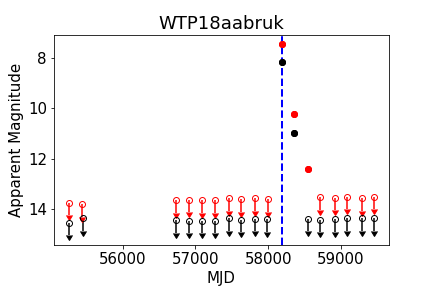}
    \includegraphics[width=0.245\textwidth]{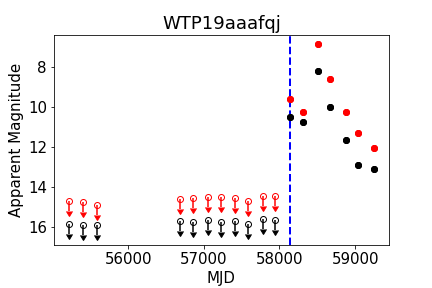}
    \includegraphics[width=0.245\textwidth]{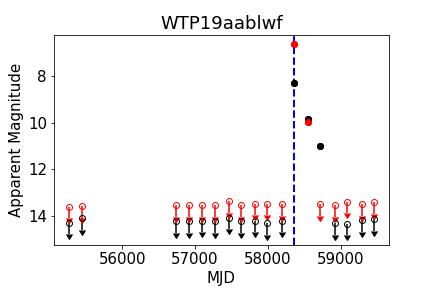}
    \includegraphics[width=0.245\textwidth]{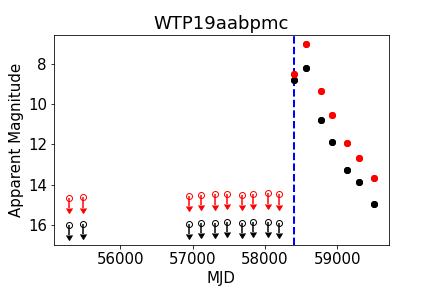}
    \includegraphics[width=0.245\textwidth]{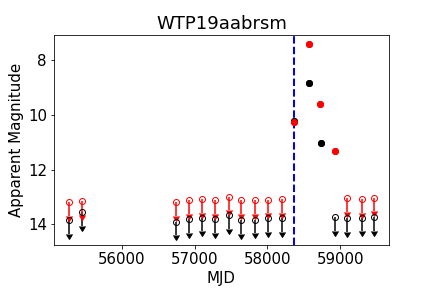}
    \includegraphics[width=0.245\textwidth]{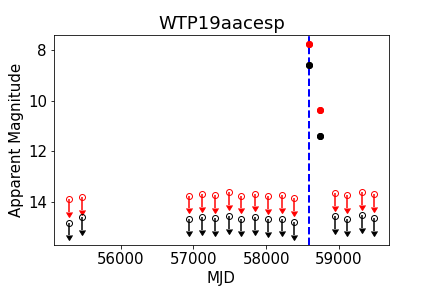}
    \includegraphics[width=0.245\textwidth]{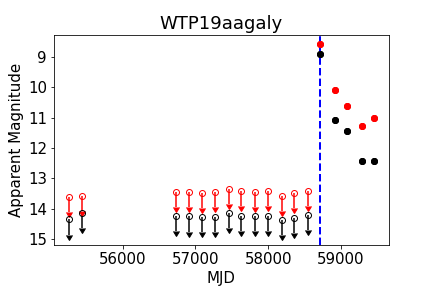}
    \includegraphics[width=0.245\textwidth]{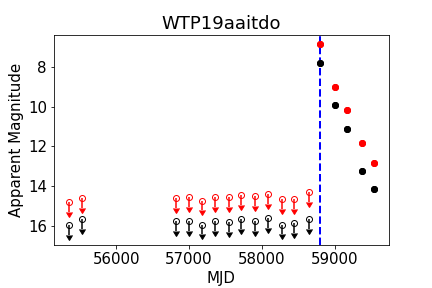}
    \includegraphics[width=0.245\textwidth]{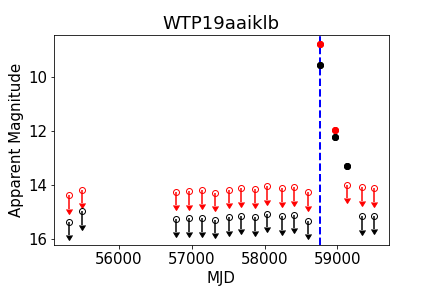}
    \includegraphics[width=0.245\textwidth]{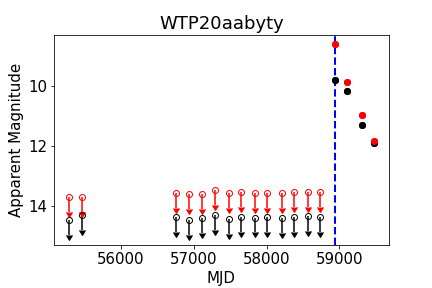}
    \includegraphics[width=0.245\textwidth]{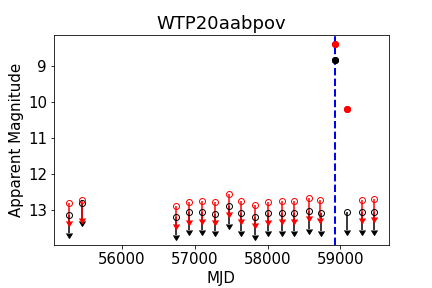}
    \caption{Difference photometry light curves for the 27 previously known nova candidates. Points plotted in black/red represent detection in the $W1$/$W2$ bands respectively. Non-detections ($SNR \leq 5$) are shown as downward facing arrows in the same colors. The first detection of the nova above the SNR threshold of 15 is marked by a blue dashed line.}
    \label{fig:known novae}
\end{figure*}

We begin our analysis by creating point spread function (PSF) forced photometry WISE light curves of all optically discovered novae in WISE over the survey period, as provided by an online database of Galactic novae\footnote{Koji’s List of Recent Galactic Novae at \url{https://asd.gsfc.nasa.gov/Koji.Mukai/novae/novae.html}.}. The resulting light curves are shown in Figure \ref{fig:known novae}. We used these light curves to define our selection criteria, which were designed to recover all similar events in our database of transient sources---including ones that were possibly missed in previous searches. We restrict our analysis to searching for outbursts that were first detected between 2015-January (to ensure at least two epochs of non-detections in the post re-activation phase of WISE) and 2020-July (in order to have at least two additional epochs of observations for the last novae in the 2021 data release). We established the following selection criteria:

\begin{enumerate}
    \item We require the transient to have peak signal-to-noise ratio (SNR) $\geq 15$, located within $|b| < 10^\circ$ and not be saturated in the images (corresponding to $W1/W2 \approx 8$\,Vega mag).
    \item The source should be at least 6\arcsec away from a known WISE source (corresponding to the approximate size of the WISE PSF; \citealt{Cutri_2012}), or if a source is present, it should be at least 3 mags brighter than any quiescent counterpart within 6\arcsec. 
    \item To remove moving foreground objects, we require that the transient have at least two detections in subsequent epochs (six months apart) satisfying our SNR threshold.
    \item We sub-select a sample of the large amplitude outbursts to identify fast rising outbursts as seen in known novae (Figure \ref{fig:known novae}), and to remove contamination from slow variable stars and slowly evolving outbursts from young stars \citep{Fischer2022}. To this end, we require that the first detection of the transient is no more than 400 days (corresponding to $\leq 2$ visits of the WISE spacecraft) before the detection that passes our threshold SNR.
    \item We visually examined the light curves and select all sources with fast fading light curves with $t_2 \leq 600$\,d and peak magnitude $ \leq 11.5$\,mag\footnote{We use magnitudes in the native WISE Vega system for the rest of this paper.} (either in the $W1$ or $W2$ bands), where $t_2$ is the time the source takes to decline by 2 magnitudes from peak. Requiring $t_2 \leq 600$\,d removes interlopers such as slowly evolving young star outbursts, while the choice of peak magnitude is motivated by the faintest expected peak brightness of a nova (absolute magnitude $M \approx -5$) within the Galaxy ($d < 20$\,kpc). We discuss the implications of excluding very faint and very slowly evolving nova outbursts ($t_2 > 600$\,d) in our search in Section \ref{sec:discussion}.
\end{enumerate}

\begin{figure*}
\centering
\small
    \includegraphics[width=0.245\textwidth]{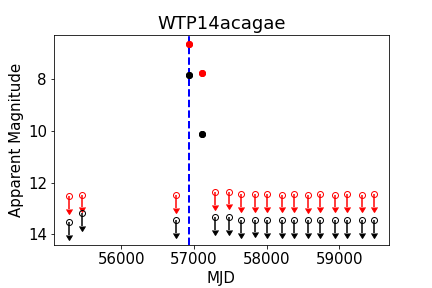}
    \includegraphics[width=0.245\textwidth]{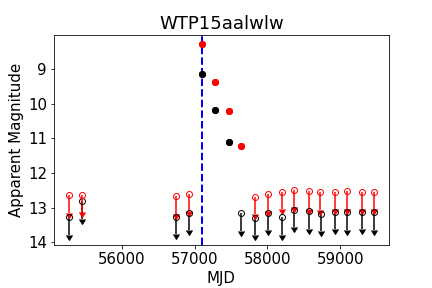}
    \includegraphics[width=0.245\textwidth]{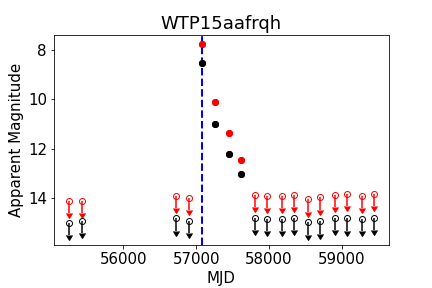}
    \includegraphics[width=0.245\textwidth]{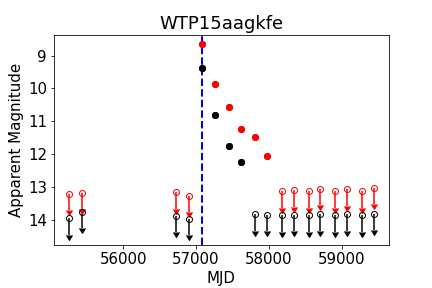}
    \includegraphics[width=0.245\textwidth]{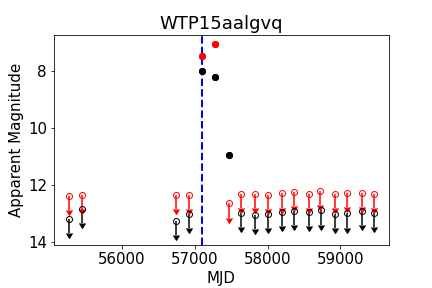}
    \includegraphics[width=0.245\textwidth]{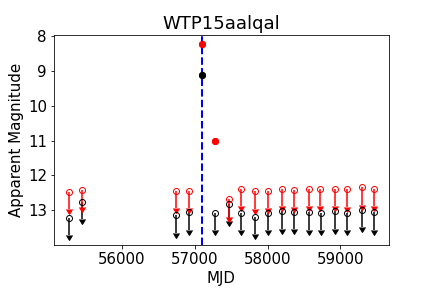}
    \includegraphics[width=0.245\textwidth]{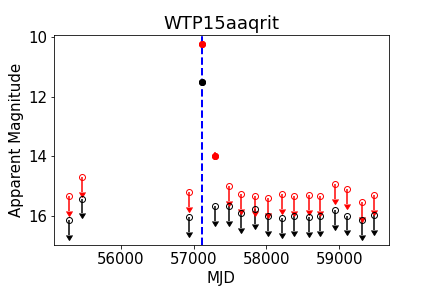}
    \includegraphics[width=0.245\textwidth]{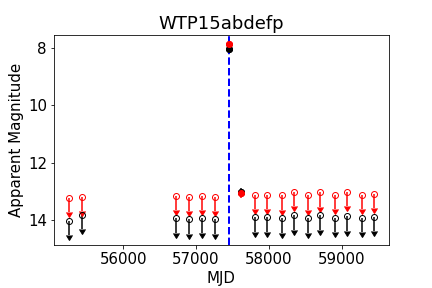}
    \includegraphics[width=0.245\textwidth]{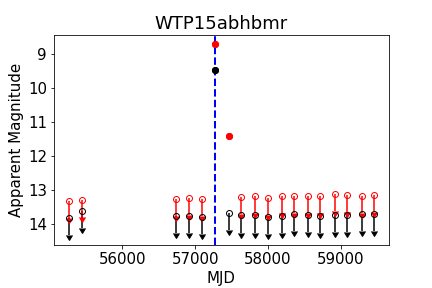}
    \includegraphics[width=0.245\textwidth]{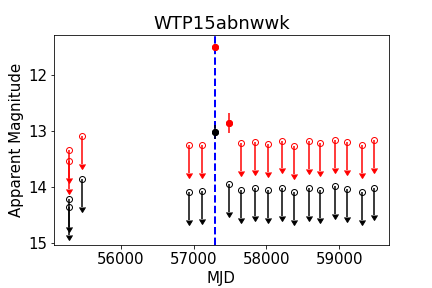}
    \includegraphics[width=0.245\textwidth]{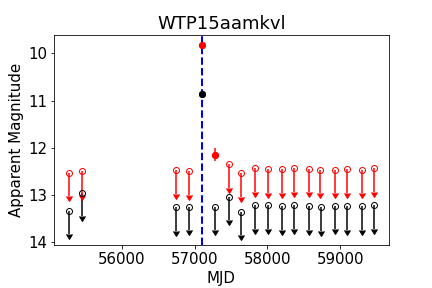}
    \includegraphics[width=0.245\textwidth]{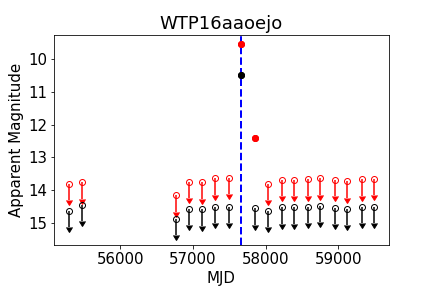}
    \includegraphics[width=0.245\textwidth]{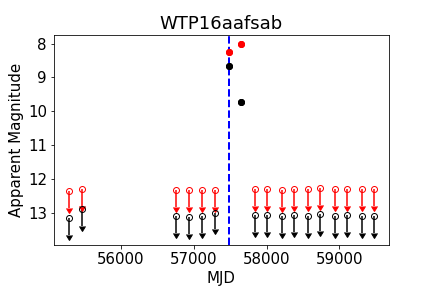}
    \includegraphics[width=0.245\textwidth]{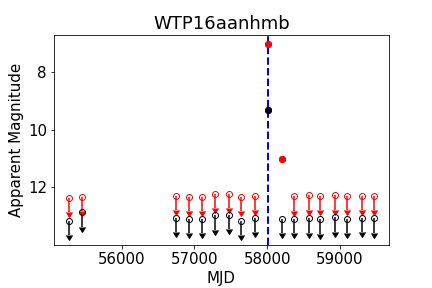}
    \includegraphics[width=0.245\textwidth]{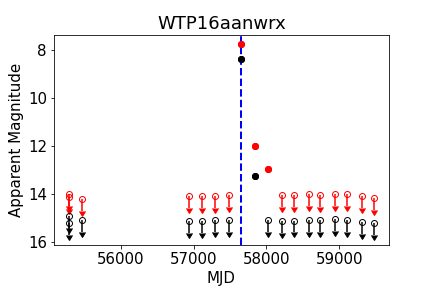}
    \includegraphics[width=0.245\textwidth]{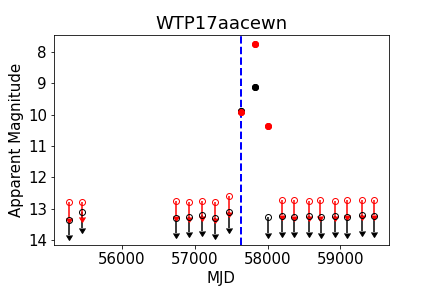}
    \includegraphics[width=0.245\textwidth]{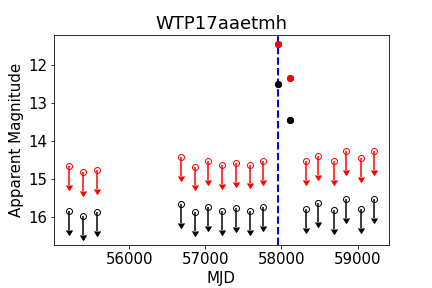}
    \includegraphics[width=0.245\textwidth]{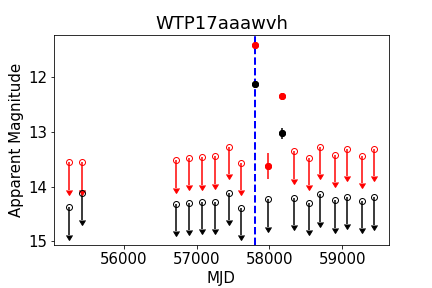}
    \includegraphics[width=0.245\textwidth]{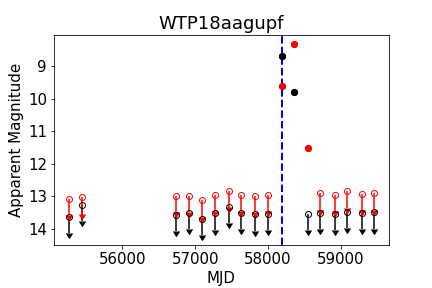}
    \includegraphics[width=0.245\textwidth]{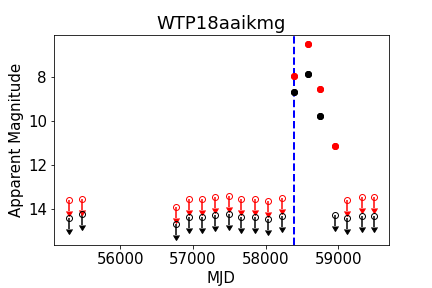}
    \includegraphics[width=0.245\textwidth]{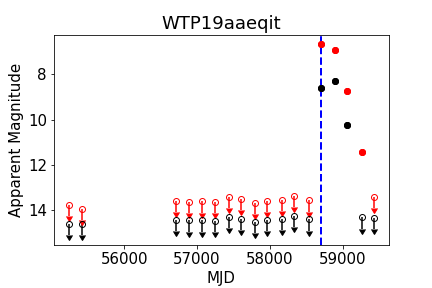}
    \includegraphics[width=0.245\textwidth]{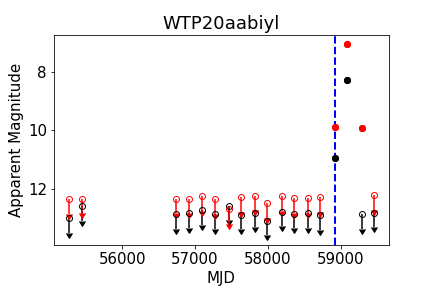}
    \caption{Difference photometry light curves for the 22 previously unknown nova candidates that have been identified in this work. Points plotted in black/red represent detection in the $W1$/$W2$ bands respectively. Non-detections ($SNR \leq 5$) are shown as downward facing arrows in the same colors. The first detection of the nova above the SNR threshold of 15 is marked by a blue dashed line.}
    \label{fig:unk novae}
\end{figure*}

Following these criteria, we rejected any remaining sources that were coincident with red mid-infrared sources in WISE to exclude likely young stellar objects \citep{Gutermuth2009}. This selection method yielded a list of 49 mid-infrared outbursts, 27 of which were previously known novae detected as transients in the WISE Transient Pipeline. The remaining 23 sources are previously unknown MIR nova candidates. In addition, the outburst of the known Galactic X-ray binary MAXI\,J1348-630 \citep{Tominaga2020} passed our selection criteria; we exclude this source from the analysis but discuss the implications of such contamination in Section \ref{sec:discussion}. The forced difference photometry light curves of the identified sources are shown in Figures \ref{fig:known novae} and \ref{fig:unk novae}. The properties of all sources detected in the pipeline are summarized in Table \ref{tab:novae}, and the corresponding ZOGY discovery image triplets are shown in Figures \ref{fig:known_nova_cutouts} and \ref{fig:unknown_nova_cutouts}.

\begin{table*}
    \footnotesize
        \centering
        \begin{tabular}{lcccccccc}
        \hline
        \hline
        Name&  Known name  & RA & Dec & Peak mag. & Amplitude &  $t_2$ & $A_V$ & $t_0$ \\ 
         & & J2000 & J2000 & Mag. (Filter) & Mag. (Filter) & days (Filter) & Mag & MJD \\
        \hline
        WTP\,19aabpmc & V435\,CMa & 07:13:45.85 & -21:12:31.43 & $7.04 \pm 0.01$ ($W2$) & $7.45 \pm 0.01$ ($W2$)& 181 ($W2$)& 2.43 & 58408.14 \\
        WTP\,18aadrrk & V549\,Vel & 08:50:29.60 & -47:45:28.55 & $7.47 \pm 0.0$ ($W2$)& $7.16 \pm 0.01$ ($W2$)&  424 ($W2$)& 3.14 & 58092.41 \\
        WTP\,19aaafqj & V357\,Mus & 11:26:14.99 & -65:31:24.18 & $6.83 \pm 0.01$ ($W2$)& $8.82 \pm 0.01$ ($W2$)& 214 ($W2$)& 2.66 & 58144.22 \\
        WTP\,17aaexxo & V1405\,Cen & 13:20:55.31  &  -63:42:19.40 & $6.87 \pm 0.01$ ($W2$) & $6.66 \pm 0.01$ ($W2$) & 98 ($W2$) & 4.55 & 57961.52\\
        WTP\,18aafjde & FM\,Cir & 13:53:27.65 & -67:25:00.62 & $6.63 \pm 0.01$ ($W2$)& $7.87 \pm 0.01$ ($W2$)&  189 ($W2$)& 1.01 & 58167.52 \\
        {\bf WTP\,19aaeqit} & - & 14:34:23.87 & -60:23:07.41 & $6.67 \pm 0.01$ ($W2$)& $7.74 \pm 0.01$ ($W2$)&  352 ($W2$)& 30.43 & 58698.28 \\
        {\bf WTP\,17aaawvh} & - & 15:10:23.73  &  -59:14:15.48 & $11.42 \pm 0.03$ ($W2$) & $2.97 \pm 0.03$ ($W2$) &  153 ($W2$) & 13.51 & 57805.54\\
        WTP\,17aaavpz & V407\,Lup & 15:29:01.79 & -44:49:40.47 & $10.55 \pm 0.01$ ($W1$)& $3.89 \pm 0.01$ ($W1$)&  122 ($W1$)& 0.88 & 57804.10 \\
        WTP\,18aafskc & V408\,Lup & 15:38:43.78 & -47:44:41.32 & $8.05 \pm 0.01$ ($W2$)& $7.49 \pm 0.01$ ($W2$)& 101 ($W2$)& 1.32 & 58334.67 \\
        {\bf WTP\,15aafrqh} & -& 15:53:53.64 & -50:32:25.59 & $7.78 \pm 0.01$ ($W2$)& $6.21 \pm 0.01$ ($W2$)&  151 ($W2$)& 5.99 & 57079.95\\
        {\bf WTP\,15abdefp} & -& 16:00:39.17 & -51:31:17.61 & $7.86 \pm 0.01$ ($W2$)& $6.12 \pm 0.01$ ($W2$)&  66 ($W2$)& 21.09 & 57447.06 \\
        {\bf WTP\,15aagkfe} & -& 16:21:44.78 & -49:02:59.04 & $8.65 \pm 0.01$ ($W2$)& $5.32 \pm 0.01$ ($W2$)& 165 ($W2$)& 24.55 & 57083.56\\
        WTP\,19aagaly & V1706\,Sco & 17:07:34.37 & -36:08:25.45 & $8.58 \pm 0.01$ ($W2$)& $4.84 \pm 0.01$ ($W2$)&  363 ($W2$)& 5.53 & 58716.56 \\
        WTP\,18aabruk & V3665\,Oph & 17:14:02.55 & -28:49:23.90 & $7.46 \pm 0.01$ ($W2$)& $6.15 \pm 0.01$ ($W2$)&  118 ($W2$)& 1.61 & 58187.93 \\
        WTP\,18aagrvf &  V1661\,Sco & 17:18:06.37 & -32:04:27.71 & $8.63 \pm 0.01$ ($W2$)& $4.73 \pm 0.01$ ($W2$)&  222 ($W2$)& 4.82 & 58188.58 \\
        WTP\,17aabqyp & V1656\,Sco & 17:22:51.43 & -31:58:36.30 & $9.61 \pm 0.01$ ($W2$) & $3.88 \pm 0.01$ ($W2$) &  136 ($W2$) & 5.36 & 57827.90 \\
        {\bf WTP\,18aagupf} &- & 17:26:36.20 & -31:02:45.05 & $8.33 \pm 0.01$ ($W2$)& $5.20 \pm 0.01$ ($W2$)&  228 ($W2$)& 8.16 & 58191.40 \\
        WTP\,15abgjhj & V2944\,Oph & 17:29:13.50 & -18:46:12.00 & $9.21 \pm 0.01$ ($W2$) & $6.36 \pm 0.01$ ($W2$) &  166 ($W2$) & 1.75 & 57271.87\\
        {\bf WTP\,15abhbmr} &- & 17:31:05.72 & -26:26:40.28 & $8.72 \pm 0.01$ ($W2$)& $5.07 \pm 0.01$ ($W2$)&  140 ($W2$)& 5.59 & 57273.71\\
         WTP\,17aabxsp & V1655\,Sco & 17:38:19.24 & -37:25:08.42 & $8.75 \pm 0.01$ ($W2$)& $5.71 \pm 0.01$ ($W2$)&  157 ($W2$)& 4.65 & 57637.52 \\
        {\bf WTP\,20aabiyl} & -& 17:38:22.26 & -31:13:21.76 & $7.07 \pm 0.01$ ($W2$)& $5.20 \pm 0.01$ ($W2$)&  256 ($W2$)& 54.79 & 58923.96 \\
        WTP\,19aablwf & V3666\,Oph& 17:42:24.08 & -20:53:08.12 & $6.64 \pm 0.01$ ($W2$)& $6.85 \pm 0.01$ ($W2$)&  121 ($W2$)& 2.35 & 58359.02 \\
        WTP\,18aahazj & Gaia18\,cew & 17:42:38.87 & -27:54:37.17 & $8.37 \pm 0.01$ ($W2$)& $4.25 \pm 0.01$ ($W2$)& 404 ($W2$)& 5.67 & 58358.36 \\
        WTP\,20aabpov & V6566\,Sgr & 17:56:13.75 & -29:42:54.60 & $8.39 \pm 0.01$ ($W2$) & $4.71 \pm 0.01$ ($W2$) &  200 ($W2$) & 3.62 & 58929.12 \\
        {\bf WTP\,15aalgvq} &- & 17:56:29.16 & -26:02:42.18 & $7.03 \pm 0.01$ ($W2$)& $5.97 \pm 0.01$ ($W2$)&  60 ($W2$)& 34.04 & 57101.42\\
        {\bf WTP\,17aacewn} &- & 17:57:59.77 & -23:18:16.49 & $7.76 \pm 0.01$ ($W2$)& $5.34 \pm 0.01$ ($W2$)&  280 ($W2$)& 14.36 & 57637.23 \\
        {\bf WTP\,15aalqal} & -& 17:59:19.75 & -24:14:32.42 & $8.24 \pm 0.01$ ($W2$)& $4.19 \pm 0.01$ ($W2$)&  128 ($W2$)& 35.70 & 57102.08\\
        WTP\,16aamwng & V5853\,Sgr & 18:01:07.74  &  -26:31:42.00 & $6.75 \pm 0.01$ ($W2$) & $6.25 \pm 0.01$ ($W2$) &  105 ($W2$) & 12.16 & 57639.98\\
        {\bf WTP\,15aalwlw} & - & 18:01:58.38 & -26:03:26.82 & $8.29 \pm 0.01$ ($W2$)& $4.86 \pm 0.01$ ($W2$)&  367 ($W2$)& 14.09 & 57102.80\\
        WTP\,16aaekyi & V5669\,Sgr & 18:03:32.76 & -28:16:04.63 & $9.82 \pm 0.01$ ($W2$)& $3.12 \pm 0.01$ ($W2$)&  229 ($W2$)& 2.41 & 57467.63 \\
        WTP\,19aabrsm & V5857\,Sgr & 18:04:09.44 & -18:03:55.98 & $7.42 \pm 0.01$ ($W2$)& $6.36 \pm 0.01$ ($W2$)&  147 ($W2$)& 7.58 & 58363.27 \\
        {\bf WTP\,15aamkvl} & - & 18:07:42.71 & -19:37:20.46 & $9.16 \pm 0.01$ ($W2$) & $4.08 \pm 0.01$ ($W2$) &  153 ($W2$) & 38.94 & 57104.44\\
        WTP\,17aahczz & V5855\,Sgr & 18:10:28.30 & -27:29:59.35 & $8.08 \pm 0.01$ ($W2$)& $5.17 \pm 0.01$ ($W2$)&  164 ($W2$)& 1.69 & 57834.35 \\
        WTP\,15aamvqj & V5667\,Sgr& 18:14:25.16 & -25:54:34.78 & $7.24 \pm 0.01$ ($W2$)& $6.06 \pm 0.01$ ($W2$)&  320 ($W2$)& 1.98 & 57105.49\\
        {\bf WTP\,16aanhmb} & -& 18:16:05.86 & -16:51:20.89 & $7.02 \pm 0.01$ ($W2$)& $6.04 \pm 0.01$ ($W2$)&  98 ($W2$)& 95.30 & 58004.80 \\
        {\bf WTP\,14acagae} &- & 18:32:43.20 & -09:08:51.12 & $6.63 \pm 0.01$ ($W2$)& $5.84 \pm 0.01$ ($W2$)&  134 ($W2$)& 57.74 & 56930.14\\
        WTP\,16aafnbt & V5668\,Sgr & 18:36:56.83 & -28:55:39.30 & $6.60 \pm 0.01$ ($W2$) & $7.53 \pm 0.01$ ($W2$) &  312 ($W2$) & 0.656 & 57110.29\\
        WTP\,20aabyty & V659\,Sct& 18:39:59.70 & -10:25:42.10 & $8.60 \pm 0.01$ ($W2$)& $5.79 \pm 0.01$ ($W2$)&  308 ($W2$)& 5.68 & 58938.80 \\
        {\bf WTP\,16aafsab} &- & 18:40:56.38 & -06:09:03.89 & $8.02 \pm 0.01$ ($W2$)& $4.97 \pm 0.01$ ($W2$)&  91 ($W2$)& 21.06 & 57477.52 \\
        {\bf WTP\,16aanwrx} &- & 18:53:07.93 & 06:44:08.82 & $7.74 \pm 0.01$ ($W2$)& $6.31 \pm 0.01$ ($W2$)&  90 ($W2$)& 9.29 & 57654.27 \\
        {\bf WTP\,15aaqrit} &- & 18:55:28.94 & 02:04:01.30 & $10.11 \pm 0.01$ ($W2$)& $3.08 \pm 0.01$ ($W2$)&  76 ($W2$)& 52.09 & 57120.01\\
        {\bf WTP\,15abnwwk} &- & 19:00:28.86 & 02:27:57.26& $11.49 \pm 0.04$ ($W2$)& $2.59 \pm 0.04$ ($W2$)& 246 ($W2$)& 18.36 & 57296.87\\
        WTP\,19aacesp & Gaia19\,buy& 19:03:14.95 & 01:20:27.91 & $7.78 \pm 0.01$ ($W2$)& $7.00 \pm 0.01$ ($W2$)&  126 ($W2$)& 9.45 & 58581.25 \\ 
        {\bf WTP\,16aaoejo} &- & 19:20:01.73 & 15:25:02.02 & $9.54 \pm 0.01$ ($W2$)& $4.98 \pm 0.01$ ($W2$)&  134 ($W2$)& 11.28 & 57666.42 \\
        WTP\,15abpbbz &V1831\,Aql & 19:21:50.16 & 15:09:25.03 & $6.60 \pm 0.01$ ($W2$)& $7.12 \pm 0.01$ ($W2$)&  171 ($W2$)& 19.29 & 57305.85\\
        {\bf WTP\,18aaikmg} &- & 19:22:34.62 & 13:01:56.91& $6.50 \pm 0.01$ ($W2$)& $7.83 \pm 0.01$ ($W2$)&  359 ($W2$)& 12.94 & 58387.49 \\
        WTP\,19aaiklb &V569\,Vul & 19:52:08.25 & 27:42:20.94 & $8.80 \pm 0.01$ ($W2$)& $6.52 \pm 0.01$ ($W2$)&  127 ($W2$)& 9.60 & 58764.85 \\
        WTP\,19aaitdo & V2891\,Cyg& 21:09:25.56 & 48:10:52.24& $6.84 \pm 0.01$ ($W2$)& $7.46 \pm 0.01$ ($W2$)&  186 ($W2$)& 12.14 & 58805.09\\
        {\bf WTP\,17aaetmh} &- & 23:10:07.48 & 61:20:23.74 & $11.44 \pm 0.01$ ($W2$)& $4.30 \pm 0.01$ ($W2$)& 142 ($W2$)& 5.14 & 57954.10\\
        \hline
        \end{tabular}
    \caption{List of all nova candidates identified in our systematic search of large amplitude, rapidly evolving outbursts in the Galactic plane with WISE. The source names highlighted in bold indicate previously unknown nova candidates.  For each source, we tabulate the celestial coordinates, the magnitude and filter of the peak outburst flux, amplitude between the pre-outburst non-detection and the peak of the outburst (in the respective filter), the signal-to-noise ratio (SNR) at peak, the estimated $t_2$ value in the filter corresponding to peak magnitude of the outburst, the integrated Galactic dust extinction along the line of sight in the visual band ($A_V$; \citealt{Schlafly2011}), and the MJD of first detection.}
    \label{tab:novae}
\end{table*}

Over the survey period considered here, 60 galactic novae were independently confirmed with spectroscopy, and 27 of these sources were identified in our search. Of the sources that were missed, five were not detected in any WISE observation due to their fast light curve evolution, while one source was observed with WISE after the search period (even though the eruption was discovered before the end of the period) and therefore did not pass our sample. 11 sources were detected as transients in the pipeline in at least one epoch, but did not pass our criteria as they faded too quickly by subsequent epochs such that they did not pass the SNR and brightness cuts (criteria (i) and (ii)). Three sources had historical variability due to prior dwarf nova outbursts or stellar variability (e.g. in symbiotic novae) than 400 days before the nova eruptions \citep{Murphy-Glaysher2022}, and therefore did not pass criterion (iv). Three sources have very slowly evolving/brightening MIR light curves with $t_2 > 600 \text{ days}$ and therefore did not pass criterion (v). Nine sources did not pass our cuts, as they were missed by the pipeline due to the imperfect recovery efficiency in dense stellar fields and near very bright stars (Section \ref{sec:rate}). We show forced difference photometry light curves of these additional 33 sources, along with detailed explanations on their behavior in Appendix \ref{sec:appendix}. We incorporate the effects of our selection criteria in our estimates of the Galactic nova rate in Section \ref{sec:rate}. Hence, none of the previously known novae detected in our pipeline over the survey period and satisfying our selection criteria were excluded from the candidate list, corroborating the consistency of our selection.

 \section{A population of missed Galactic nova candidates}
 \label{sec:characteristics}

 \begin{figure*}
    \centering
    \includegraphics[width=\textwidth]{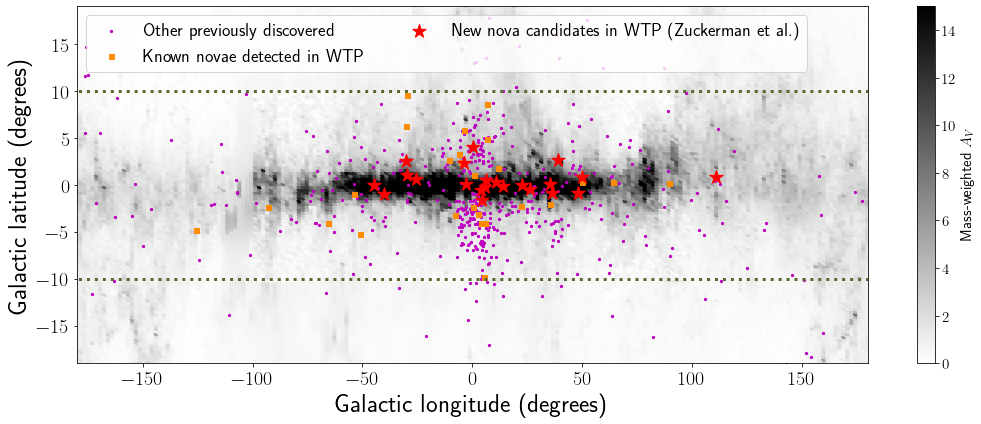}
    \caption{Sky distribution of all known novae (purple dots), known novae recovered in our WISE search (orange) and new nova candidates (red) identified in this work. The background shows a grayscale map of the mass-weighted extinction along each line of sight, using the Galactic mass model of \citet{Cautun2020} and the 3D Galactic dust model of \citet{Bovy2016}. The dotted green lines show the boundaries of the sky region where we searched for outbursts.}
    \label{fig:nova_sky}
\end{figure*}

Our search has identified a population of rapidly evolving mid-infrared outbursts near the Galactic plane in WISE data---nearly half of which are previously unknown. Figure \ref{fig:unk novae} shows that the general light curve morphologies of the newly identified sources bear striking resemblance to those of previously discovered novae (Figure \ref{fig:known novae} and Appendix \ref{sec:appendix}). However, the discussion in the previous section and comparison to the light curves in Appendix \ref{sec:appendix} shows that this sample is not sensitive to fast novae (e.g. \citealt{Aydi2018}) that are not detected at high significance in two subsequent epochs---assuming a typical peak WISE magnitude of $\approx 7$\,mag (corresponding to a $M\approx -7.5$ nova at the Galactic center) and a single epoch $15\sigma$ depth of $\approx 14$\,mag, we estimate that our search is roughly sensitive to nova outbursts with $t_2 \gtrsim 50$\,days, consistent with the properties of the identified events (Table \ref{tab:novae}). At the same time, our selection excludes very slowly evolving novae in order to remove contamination from other stellar outbursts, and therefore is not sensitive to very slowly evolving novae (e.g. \citealt{Munari2022}). Using published distributions for the decline times of novae \citep{Kawash2021b}, we therefore estimate a strict upper limit for the recovery fraction of Galactic novae in WISE of $\lesssim 20$\% based on timescales .

Figure \ref{fig:nova_sky} shows the sky positions of novae identified in WISE data (in Galactic coordinates), together with all previously discovered events. In order to assess the sensitivity of this mid-infrared search to dust-obscured events, we follow the procedure laid out in \citet{De_2021} to estimate the mass-weighted dust extinction along each line of sight. We use the Galactic stellar mass distribution from \citet[their Tables 1 and 2]{Cautun2020} for a contracted halo model, together with the three-dimensional Galactic extinction map \texttt{Combined19} from \citet{Bovy2016}. If novae approximately follow the mass distribution of the Milky Way, then the resulting dust map (shown in Figure \ref{fig:nova_sky}) traces the expected visual extinction to novae along the same line of sight. As shown in Figure \ref{fig:nova_sky} and previously noted by \citet{De_2021}, previously discovered samples of optical novae are severely biased toward finding novae away from the central Galactic plane---which has the highest mass density (and resulting highest expected nova rate) and highest dust extinction in the Galaxy. However, the WISE sample of novae, being discovered in the mid-infrared bands over a broad-sky area including the Galactic plane, is tightly clustered around the central Galactic plane in regions of heavy extinction. This trend corroborates previous suggestions \citep{De_2021, Kawash_2022} that infrared searches are substantially more sensitive to the (most frequent) nova eruptions near the Galactic plane, even though these events have been historically overlooked in optical surveys.

\section{The Galactic nova rate}
\label{sec:rate}

\begin{figure*}
    \centering
    \includegraphics[scale = 0.40]{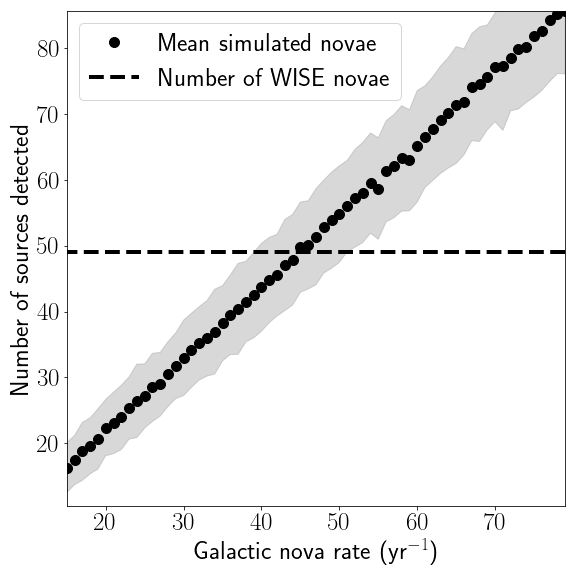}
    \includegraphics[scale = 0.40]{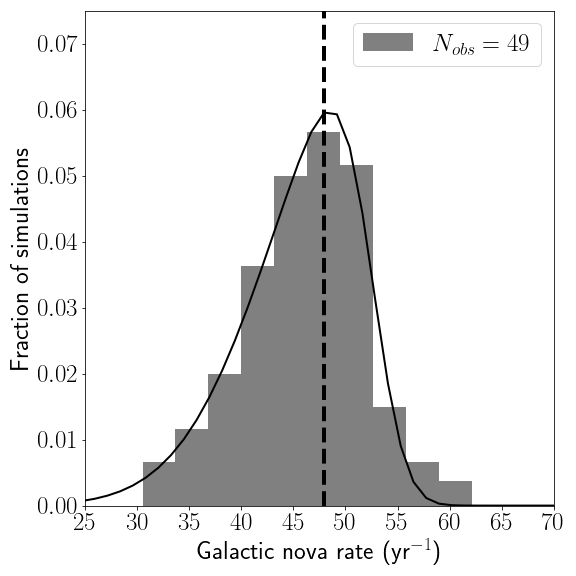}
    \caption{ (Left) Simulated number of novae recovered using our WISE pipeline and selection criteria as a function of the input nova rate. The actual number of nova candidates identified with our selection is shown by the dashed line. (Right) Distribution of the fraction of simulations that produce the exact number of novae detected by WISE (shown in the legend) as a function of the input nova rate. The solid black curve represents the skewed Gaussian distribution fit, and the dashed line represents the best fit nova rate derived from the maximum of the fitted curve.}
    \label{fig:nova sim plot}
\end{figure*}

The abundance of luminous, yet often overlooked, mid-infrared outbursts in the Galactic plane from the last decade demonstrates the efficacy of our search method in finding missed novae in our Galactic backyard. In order to quantify the implications of these discoveries in the context of the Galactic nova rate, we carry our detailed simulations of the WISE survey strategy and the detection efficiency of the image subtraction and candidate selection methods. We follow the framework laid down by \citet{De_2021} to carry out this procedure. In brief, we first estimate the detection efficiency for transients in our subtraction pipeline over a set of $144$ fields within 10 degrees of the Galactic plane (the search region for our selection) and spread over the full range of Galactic longitudes. For each field, we inject fake transients into the images over a range of magnitudes in the dynamic range of the WISE images, and run the images through our subtraction pipeline to compute the detection efficiency. The resulting efficiencies range from $\approx 20-30$\% for fields at the Galactic center (where stellar confusion noise severely hampers transient discovery) to $\approx 95-100$\% along most of the Galactic plane ($|l| \gtrsim 10^\circ$. Further details will be presented in De et al. in prep.

We then ran Monte Carlo simulations of the WISE survey strategy by assuming a population of novae erupting over the search period following the stellar mass distribution of the Milky Way \citep{Cautun2020}. We do not include the effects of dust extinction as our search is in the mid-infrared bands where extinction is $\lesssim 5$\% of the optical bands \citep{Cardelli1989}. For each simulated survey, we adopt an input rate, $r_0$, and a fixed duration of the survey, $t_s$, and create a population of novae with the total number of events following a Poisson distribution (given below) with a mean of $\lambda = r_0 * t_s$. 
\begin{equation}
    P(N, \lambda) = \frac{\lambda^N}{N!}e^{-\lambda}
\end{equation}

For each simulated nova in the population, we adopt a peak absolute magnitude from a Gaussian distribution with mean $\mu = -7.2$ and standard deviation $\sigma = 0.8$ (in the Vega magnitude system), based on the luminosity function of novae in M31 \citep{Shafter_2017}. In doing so, we assume that the mid-infrared luminosity function of novae (in Vega magnitudes) follows the optical luminosity function, which holds true if novae have colors similar to A0V stars at peak light (e.g. \citealt{Shafter2009}). However, we caution that novae forming dust at late times may be more luminous and long-lived than our nominal assumptions (e.g. \citealt{Kasliwal2017}). The absolute magnitude sets the evolution timescale of the nova $t_2$, which was calculated from the Maximum Magnitude Relation with Decline time (MMRD; \citealt{Zwicky_1936, Mclaughlin_1945}) in \citet{ozdonmez}. For the entire population of simulated novae, we evaluated whether each object would pass the selection criteria laid out in Section \ref{sec:identification} assuming novae follow a linear decline from peak magnitude set by the decline timescale $t_2$.

For each assumed Galactic nova rate, we ran 200 simulations for the nova population, in each case computing the number of events that would pass our selection criteria. We show the results in Figure \ref{fig:nova sim plot}, demonstrating how the recovered number of novae increases with the assumed Galactic rate. In order to estimate the best-fit Galactic rate corresponding to the number of events actually identified, we follow \citet{De_2021} and compute the fraction of simulations that produce the observed number of events (=49) as a function of the input Galactic rate. We fit a skewed Gaussian function to this distribution to estimate the best-fit rate and its confidence intervals. We thus derive a rate and its 68\% confidence interval
\begin{equation}
\centering
    r_0 = 47.9^{+3.1}_{-8.3} \text{yr}^{-1}
\end{equation}
The derived rate corresponds to our assumption of a uniform specific rate of novae in the bulge and disk populations, and for our assumed luminosity function. We discuss the implications of our selection criteria and model assumptions in Section \ref{sec:discussion}.

\section{Discussion}
\label{sec:discussion}

We have thus far presented a systematic search for rapidly evolving Galactic plane outbursts in NEOWISE mid-infrared data, identifying 49 transients as Galactic nova candidtaes, out of which 27 are previously confirmed novae. The remaining sources constitute previously unknown nova candidates. Using detailed simulations to estimate the effects of the WISE survey strategy, the pipeline recovery efficiency, and our selection criteria, we estimate the Galactic nova rate implied our observations. In this section, we provide more detailed discussions on the detection efficiency for different nova populations as well as limitations posed by our identification criteria. 

We first consider our confidence in the nature of the identified outbursts. As this work focuses on an archival search for events, spectroscopic confirmation of the events remains highly unfeasible. Therefore, although our search identifies more events than other recent systematic searches \citep{De_2021, Kawash_2022}, there is likely contamination from other fast evolving outbursts (e.g. X-ray binaries and young stars) near the Galactic plane (noting that the sample of \citealt{Mr_z_2015} as well as \citealt{Kawash_2022} also consist of several unconfirmed events). The identification of the mid-infrared outburst of the X-ray binary MAXI\,J1348-630 with our selection criteria further corroborates this point; to this end, we checked the online\footnote{\url{http://xmmuls.esac.esa.int/upperlimitserver/}} High Energy Upper Limit Server \citep{Saxton2022} to search for known X-ray emission coincident with any of the new candidates, and did not find any convincing detections. Similarly, large amplitude young star outbursts (e.g. the FU Ori class of events) are known to emerge as extremely long and slowly evolving ($\approx$ years - decades) mid-infrared outbursts (e.g. \citealt{Hillenbrand2018}) and are not included with our selection criteria. Faster-evolving outbursts of the EXor class may contaminate this sample, noting however that the $> 10$\,year WISE baseline and the history of inactivity in all the previously unknown sources argues against the young star hypothesis (which are intrinsically highly variable sources even in quiescence; \citealt{Bonito2018}). None of the sources are expected to be dwarf novae given their long timescales (which contaminate faster cadence ground-based searches; \citealt{Kawash_2021, De_2021}). 

Next, we discuss the effects of our model assumptions on the nova distribution in the Milky Way. As an all-sky survey, our WISE search is sensitive to novae erupting along the entire Galactic bulge and disk regions. However, the slow survey cadence and lower recovery efficiency near the Galactic center severely impact the recovery of events. Using our survey simulations, we estimate that our selection recovers $\approx 20$\% of all Galactic novae that erupted over the simulation period, consistent with our previous rough estimate. Separating the novae into bulge and disk populations, we find the recovery efficiency to be $\approx 16$\% and $\approx 21$\% for bulge and disk novae, respectively, with the lower bulge rate arising due to the lower transient recovery efficiency in dense stellar fields. If we instead assume a brighter luminosity function of $M = -7.9 \pm 0.8$ as suggested for Milky Way novae (\citealt{ozdonmez}; see however arguments against this conclusion in \citealt{Shafter_2017}), the corresponding recovery efficiencies drop to $\approx 13$\%, $\approx 9$\% and $\approx 11$\% for disk, bulge and all novae respectively. In particular, the faster light curve evolution implied by a brighter nova population decreases the fraction of novae that would be recovered in our slow cadence WISE search.

As all the previously unknown events (constituting $>40$\% of the total sample) do not have spectroscopic confirmation, considering the effects of differing luminosity functions (e.g. between bulge and disk novae), Galactic structures, and populations of novae possibly differing from the MMRD (e.g. \citealt{Kasliwal2011}) is beyond the scope of the paper. While it is possible to then consider the derived estimate as an upper limit to the Galactic nova rate, we caution that our search is not sensitive to slowly evolving eruptions from novae forming dust at late times (e.g. \citealt{Gehrz1995}; Appendix \ref{sec:appendix}) or to events that have prior history of variability in dwarf novae \citep{Murphy-Glaysher2022} or symbiotic systems \citep{Schaefer2010}. Certain sub-types of novae forming dust at late times, or displaying MIR coronal emission lines are indeed known to have long-lived IR light curves (\citealt{Gehrz1995}; see also Figure \ref{fig:known novae}) which may be missed by our search due to their extremely slow IR evolution. Comparing our rate estimate to recent works, we find it to be consistent with the recent estimates of \citet{Shafter_2017}, \citet{De_2021} and (marginally) with \citet{ozdonmez}. Our estimates are nominally inconsistent with the lower recent estimates of \citet{Kawash_2022}, who used optically discovered novae; however, the lack of complete spectroscopic confirmation in both our and the \citet{Kawash_2022} sample limits the interpretation.

\section{Summary}
\label{sec:summary}

In this paper, we utilize the long and uniform baseline of the NEOWISE survey to identify a systematic sample of large-amplitude, rapidly evolving outbursts near the Galactic plane ($|b| < 10^\circ$) in search of slowly evolving Galactic novae. The mid-IR sensitivity of NEOWISE at $3.4$ and $4.6$\,$\mu$m (where interstellar extinction is minimal) offers the opportunity to uncover luminous stellar eruptions in the most dust-obscured regions of the Milky Way, where both ground-based optical and NIR surveys become insensitive, thereby providing an independent method to probe the demographics of novae. We define a set of criteria motivated by the observed mid-IR evolution of previously confirmed novae, and we find a sample of 49 Galactic plane outbursts. While 27 of these outbursts are previously confirmed novae, the remaining sources constitute a sample of bright Galactic MIR outbursts reported for the first time in this work. The MIR light curves of the previously unknown sources bear striking resemblance to confirmed novae, which, together with their large peak bright brightness, makes them strong candidates for novae overlooked in our Galactic backyard. By using a a luminosity function following that of M31 novae \citep{Shafter_2017} and simulating the WISE survey strategy, as well our selection criteria for a fiducial population of novae tracing the Galactic stellar--mass distribution, we perform Monte Carlo simulations to estimate the Galactic nova rate to be $47.9^{+3.1}_{-8.3}$\,yr$^{-1}$.

Given that spectroscopic confirmation remains difficult for these archival sources due to their rapid fading post outburst peak, we discuss how our selection strategy possibly includes other types of Galactic stellar outbursts (e.g. X-ray binaries, young stars) and excludes rarer types of novae with pre-outburst variability and slow late-time infrared evolution due to dust formation or coronal line emission. Some of these interlopers may be confirmed or rejected by crossmatching this sample with upcoming multi-wavelength, time-domain survey catalogs in the X-ray (e.g. eROSITA; \citealt{Predehl2021}) and radio (e.g. Very Large Array Sky Survey; \citealt{Lacy2020}) bands. Nevertheless, the discovery of such a large sample of previously unknown outbursts in the mid-infrared highlights how prior optical searches have been severely limited in finding bright stellar eruptions in our  Galactic neighborhood (Figure \ref{fig:nova_sky}). In particular, this systematic archival search highlights the potential for real-time search and characterization of MIR transients if NEOWISE images are made available to the community shortly after data acquisition (compared to the current yearly data releases) and in the future, with the NEO Surveyor mission \citep{Mainzer2022, 2019BAAS...51c.321R, 2019BAAS...51c.108K}. 

\section*{Data Availability}
The data presented here are based on an analysis of publicly available images from the \texttt{unwise} project\footnote{\url{http://unwise.me/}}. The WISE light curves of the nova candidates identified here, as well as previously known events that did not pass our thresholds (Appendix \ref{sec:appendix}) will be provided along with the publication.

\section*{Acknowledgements}
We thank E. Kara, R. Simcoe and M. M. Kasliwal for helpful discussions on the work. L. Z. would like to thank the Research Science Institute, the Center for Excellence in education, and all associated sponsors for supporting this research. K. D. was supported by NASA through the NASA Hubble Fellowship grant \#HST-HF2-51477.001 awarded by the Space Telescope Science Institute, which is operated by the Association of Universities for Research in Astronomy, Inc., for NASA, under contract NAS5-26555. This publication makes use of data products from the Wide-field Infrared Survey Explorer, which is a joint project of the University of California, Los Angeles, and the Jet Propulsion Laboratory/California Institute of Technology, funded by the National Aeronautics and Space Administration. 

\bibliographystyle{mnras}

%\bibliography{example} % if your bibtex file is called example.bib

%%%%%%%%%%%%%%%%%%%% REFERENCES %%%%%%%%%%%%%%%%%%

% The best way to enter references is to use BibTeX:

% Alternatively you could enter them by hand, like this:
% This method is tedious and prone to error if you have lots of references
%\begin{thebibliography}{99}
%\bibitem[\protect\citealtauthoryear{Author}{2012}]{Author2012}
%Author A.~N., 2013, Journal of Improbable Astronomy, 1, 1
%\bibitem[\protect\citealtauthoryear{Others}{2013}]{Others2013}
%Others S., 2012, Journal of Interesting Stuff, 17, 198
%\end{thebibliography}

%%%%%%%%%%%%%%%%%%%%%%%%%%%%%%%%%%%%%%%%%%%%%%%%%%

%%%%%%%%%%%%%%%%% APPENDICES %%%%%%%%%%%%%%%%%%%%%

\include{appa}

%%%%%%%%%%%%%%%%%%%%%%%%%%%%%%%%%%%%%%%%%%%%%%%%%%
\let\clearpage\relax

% Don't change these lines
\bsp	% typesetting comment
\label{lastpage}
\end{document}

%% file: appa.tex
\appendix
\onecolumn

\section{Discovery image cutouts at the locations of the identified nova candidates}
\begin{figure}
    \centering
    \includegraphics[width=\textwidth]{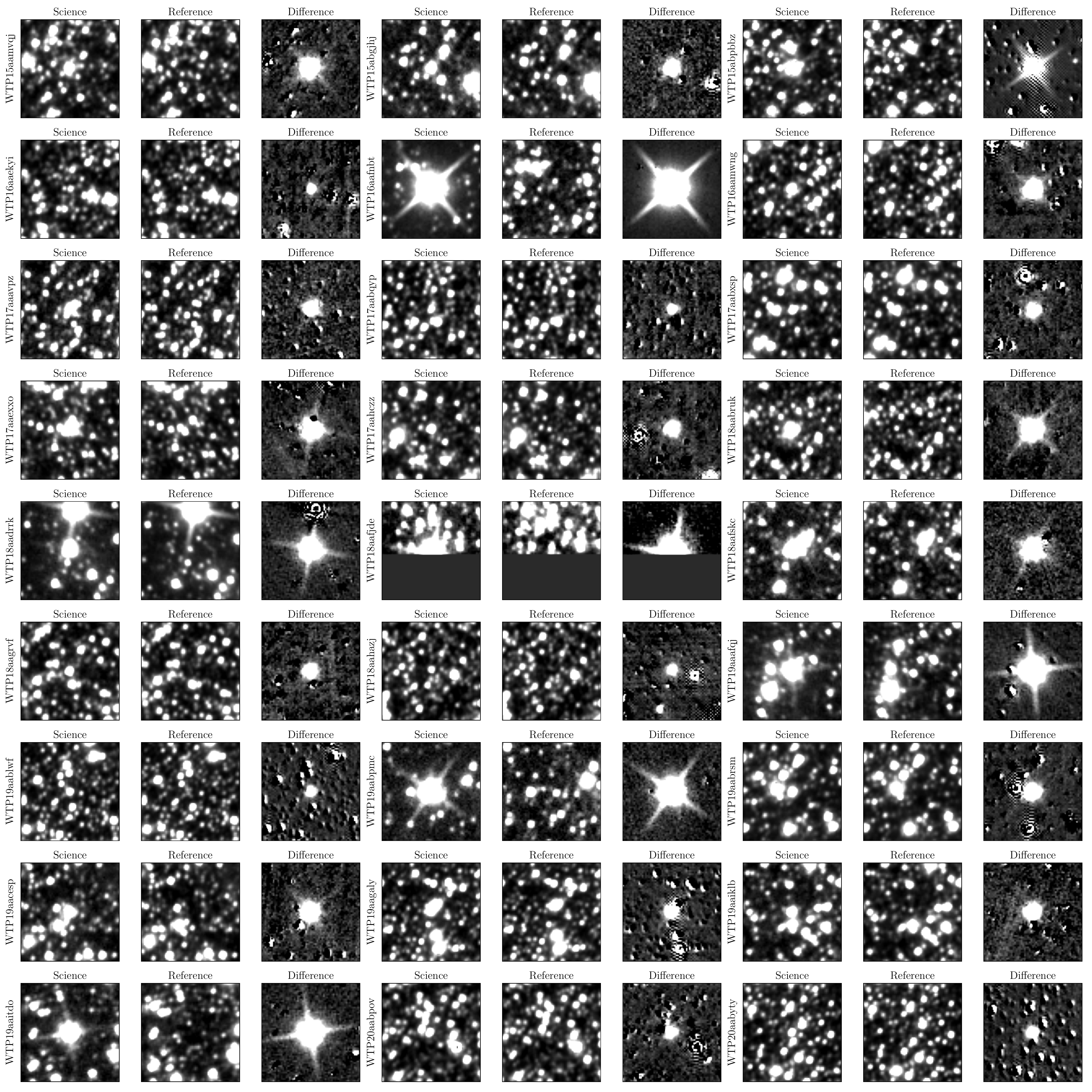}
    \caption{Discovery triplets of all previously known novae in the identified sample. We show science, template and difference images for each source (as indicated on the top labels), with the nova WTP name indicated on the left of each triplet.}
    \label{fig:known_nova_cutouts}
\end{figure}

\begin{figure}
    \centering
    \includegraphics[width=\textwidth]{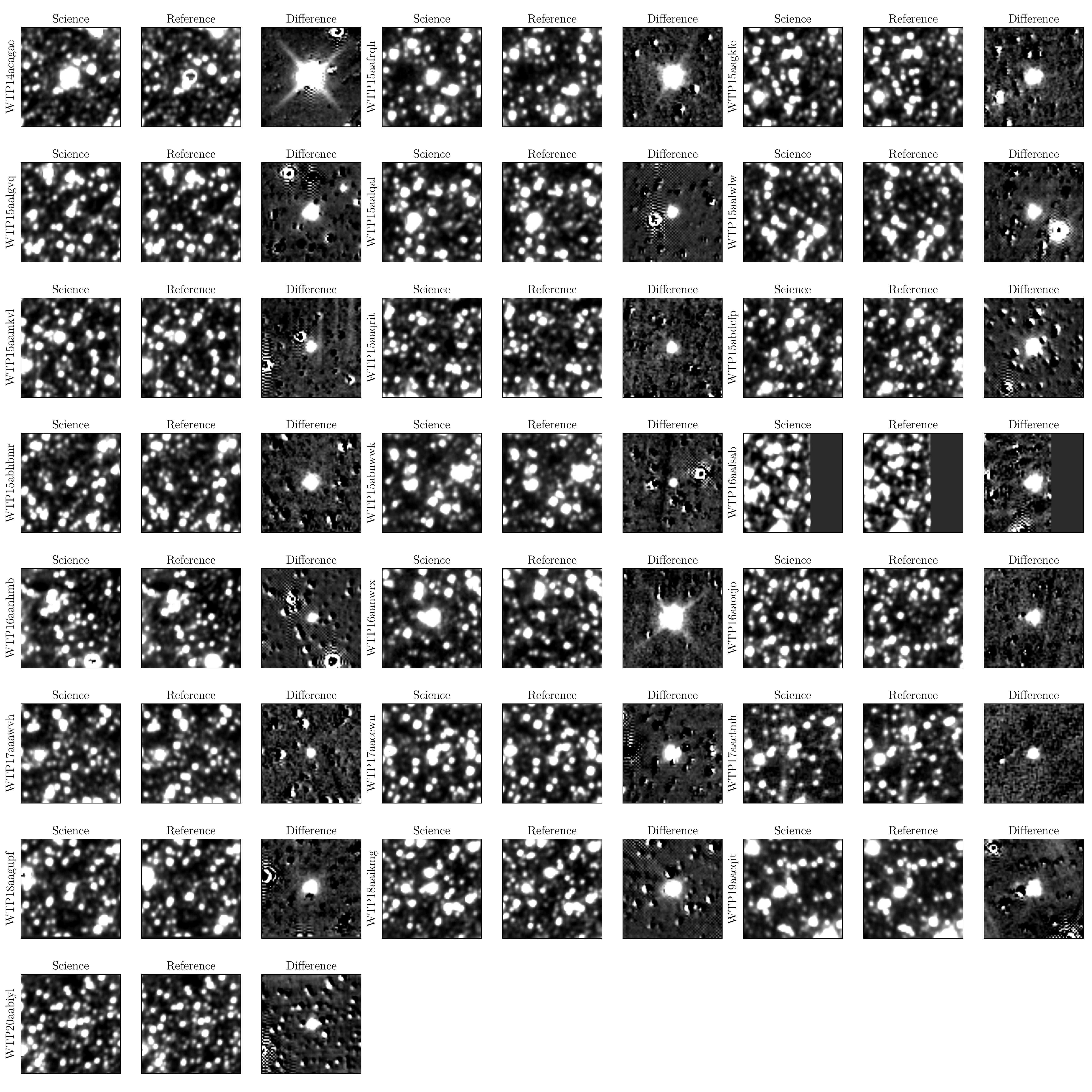}
    \caption{Same as Figure \ref{fig:known_nova_cutouts} for the previously unknown nova candidates identified in this work.}
    \label{fig:unknown_nova_cutouts}
\end{figure}

In Figures \ref{fig:known_nova_cutouts} and \ref{fig:unknown_nova_cutouts}, we show the science, template and difference image triplets for all the novae that passed our thresholds. For each source, we show the epoch corresponding to the epoch of brightest detection in the $W1$/$W2$ band. Several sources are close to the nominal saturation threshold of WISE at their peak brightness, and the transient sources show clear artifacts from diffraction spikes. 

\section{WISE light curves of known novae that were excluded in the search}
\label{sec:appendix}
Given the selection criteria and identification methods in our WISE search, a number of previously confirmed novae did not pass our thresholds and were therefore excluded from our demographic analysis. However, a number of these sources have clear detections in WISE data when performing forced photometry at the known sky locations. In this section, we provide for completeness, WISE $W1$/$W2$ light curves of all these novae and briefly discuss their properties. As mentioned in the main text, the excluded novae constitute a combination of fast fading events that were not detected in multiple visits of WISE, events that had prior history of variability and were excluded or were not recovered due to the inefficiencies of the subtraction pipeline. First, the sources V1660\,Sco, V1707\,Sco, V3730\,Oph, V3663\,Oph and V5850\,Sgr were not detected in a single epoch of WISE forced photometry, suggesting that they faded too quickly to be included in the sample. Although V6567\,Sgr erupted within the search period, the first WISE observations of the source were acquired in 2021 and therefore excluded from our sample.  Next, the sources ASASSN\,17no, V1658\,Sco, V2860\,Ori, V3661\,Oph, V5854\,Sgr, V611\,Sct, V613\,Sct, V670\,Ser, V1657\,Sco, V556\,Nor and Gaia\,20dfb are faster novae that were detected in at least one epoch, but faded too quickly by the time of the second epoch such that the SNR threshold or magnitude cut requirement was not satisfied at the second epoch (though the transient many have been formally detected in forced photometry). V1535\,Sco, V3890\,Sgr and V3664\,Oph are symbiotic novae coincident with bright evolved stars, and have variability prior to the nova outburst and therefore were excluded by the selection. V392\,Per exhibited a dwarf nova outburst several years prior to the nova and did not pass our selection. The slowly evolving / brightening (instead of fading) MIR light curves of V1709\,Sco, V5856\,Sgr and V906\,Car were excluded from our selection for fast fading outbursts. The sources V612\,Sct, V555\,Nor, V3662\,Oph, V2000\,Aql, and V1662\,Sco have light curves in forced photometry that would formally pass our cuts; however the imperfect detection efficiency of our pipeline led to incomplete recovery of the transient in some epochs, leading to these sources being excluded from our search. The sources V3731\,Oph, V2949\,Oph, V1663\,Sco and V1659\,Sco were masked out in the pipeline due to their proximity to a very bright star ($< 8$\,mag) in the template WISE images. We attempt to incorporate the last two effects with our simulations of the pipeline efficiency in Section \ref{sec:rate}.

\begin{figure}
    \small
    \centering
    \includegraphics[width=0.245\textwidth]{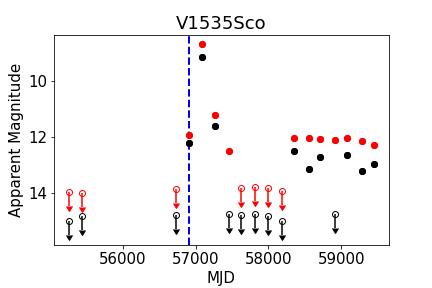}
    \includegraphics[width=0.245\textwidth]{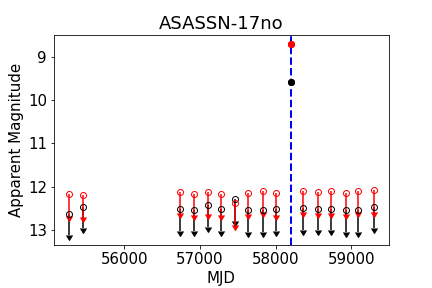}
    \includegraphics[width=0.245\textwidth]{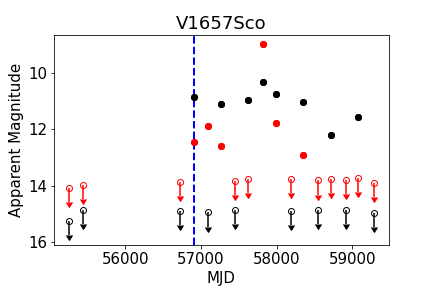}
    \includegraphics[width=0.245\textwidth]{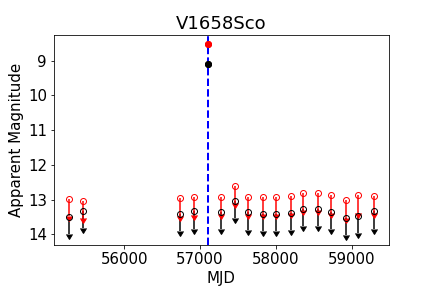}
    \includegraphics[width=0.245\textwidth]{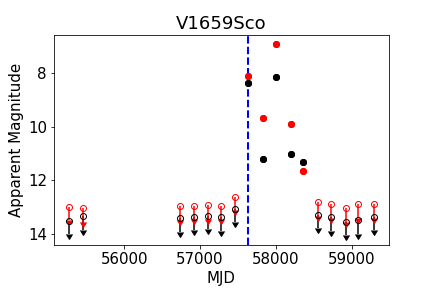}
    \includegraphics[width=0.245\textwidth]{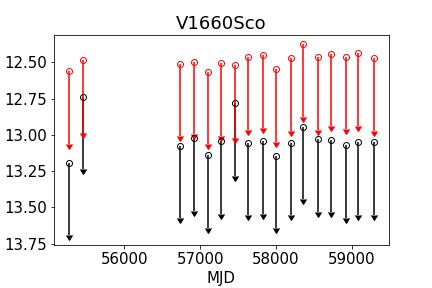}
    \includegraphics[width=0.245\textwidth]{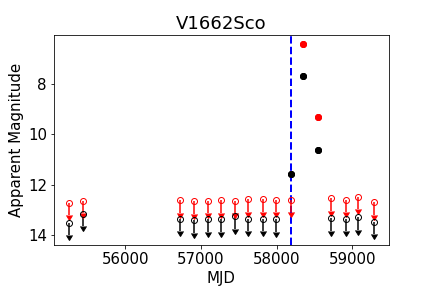}
    \includegraphics[width=0.245\textwidth]{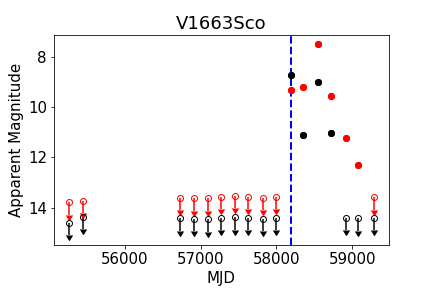}
    \includegraphics[width=0.245\textwidth]{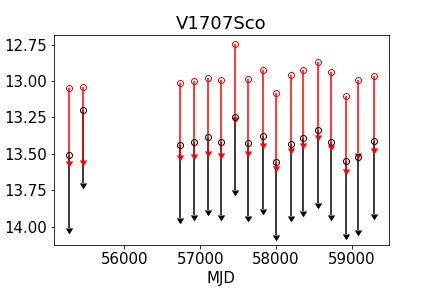}
    \includegraphics[width=0.245\textwidth]{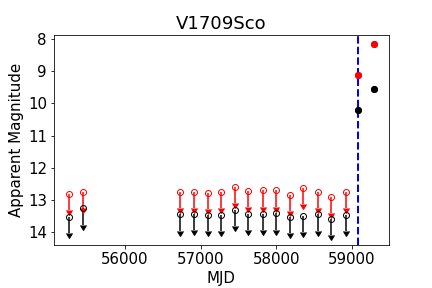}
    \includegraphics[width=0.245\textwidth]{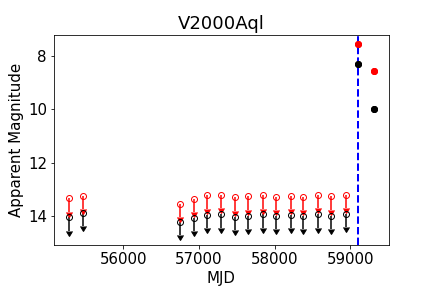}
    \includegraphics[width=0.245\textwidth]{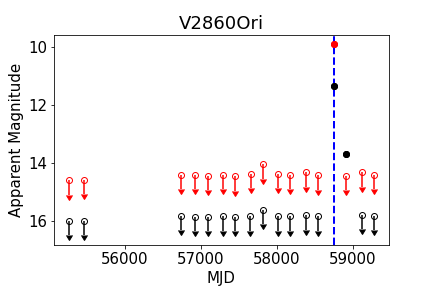}
    \includegraphics[width=0.245\textwidth]{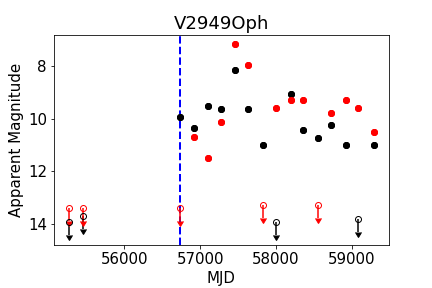}
    \includegraphics[width=0.245\textwidth]{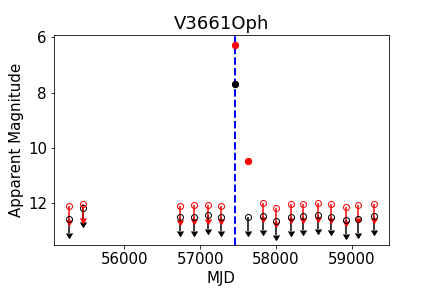}
    \includegraphics[width=0.245\textwidth]{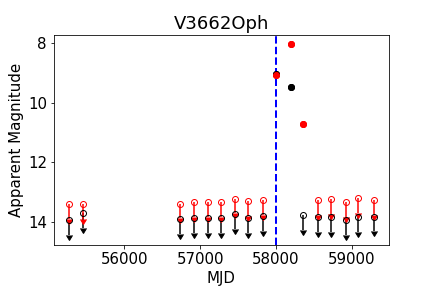}
    \includegraphics[width=0.245\textwidth]{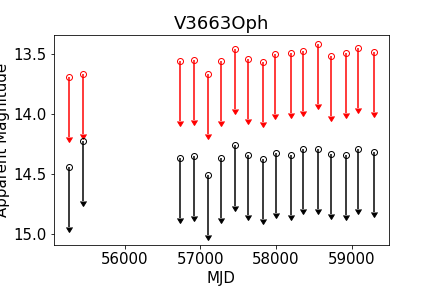}
   \includegraphics[width=0.245\textwidth]{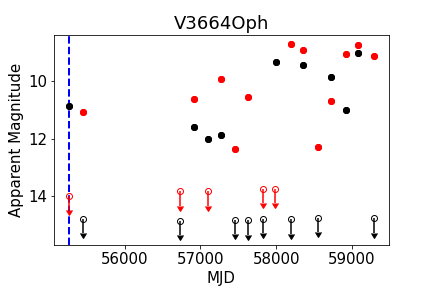}
    \includegraphics[width=0.245\textwidth]{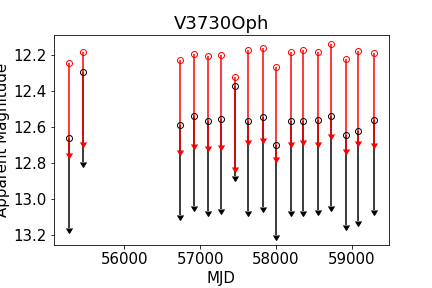}
    \includegraphics[width=0.245\textwidth]{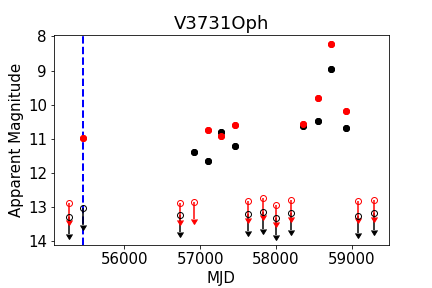}
    \includegraphics[width=0.245\textwidth]{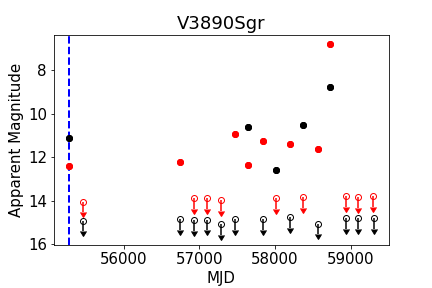}
    \includegraphics[width=0.245\textwidth]{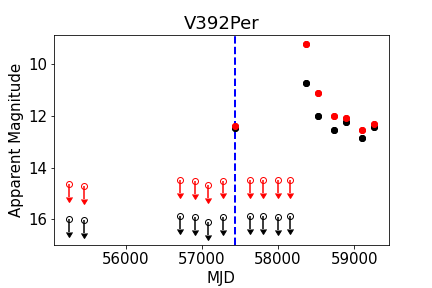}
    \includegraphics[width=0.245\textwidth]{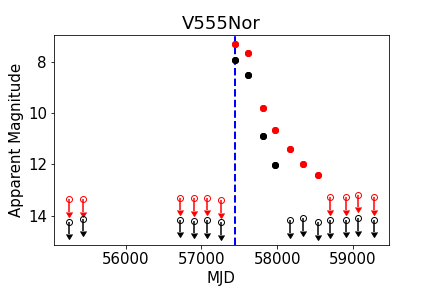}
    \includegraphics[width=0.245\textwidth]{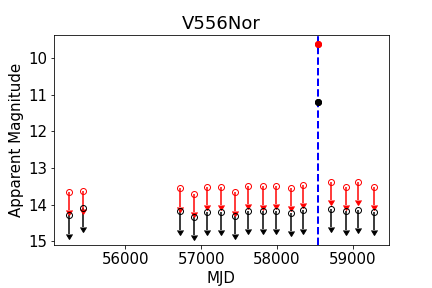}
    \includegraphics[width=0.245\textwidth]{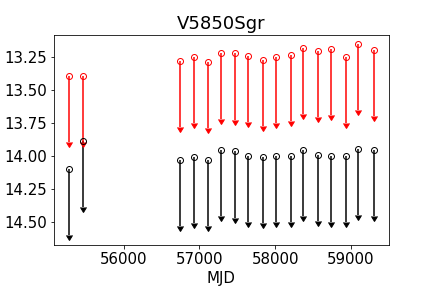}
    \caption{WISE forced difference photometry light curves of all known Galactic novae within our search period that did not pass our selection criteria. The first detection of the nova above the SNR threshold of 15 is marked by a blue dashed line. In several cases, the history of activity prior to the main outburst is due to artifacts introduced by the presence of nearby bright stars or due to intrinsic variability from the donors in symbiotic novae and previous dwarf nova outbursts.}
    \label{fig:add_lcs}
\end{figure}

\begin{figure}
\ContinuedFloat
    \centering
    \small
    \includegraphics[width=0.245\textwidth]{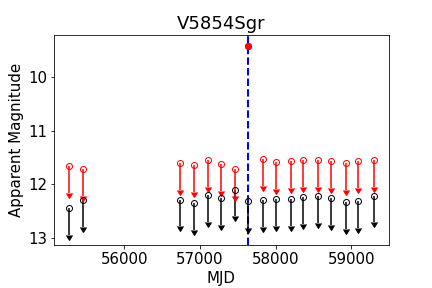}
      \includegraphics[width=0.245\textwidth]{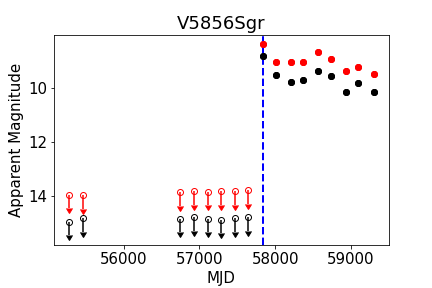}
    \includegraphics[width=0.245\textwidth]{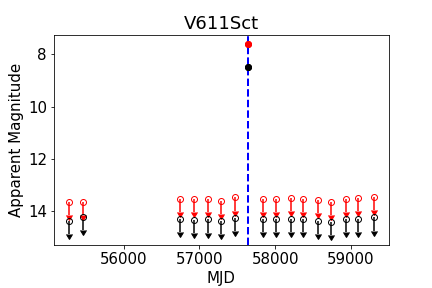}
    \includegraphics[width=0.245\textwidth]{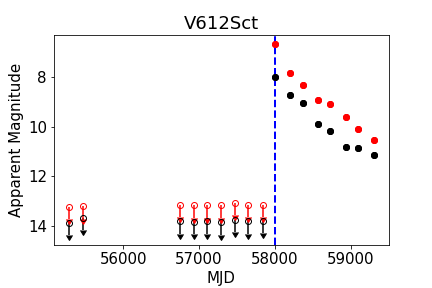}
    \includegraphics[width=0.245\textwidth]{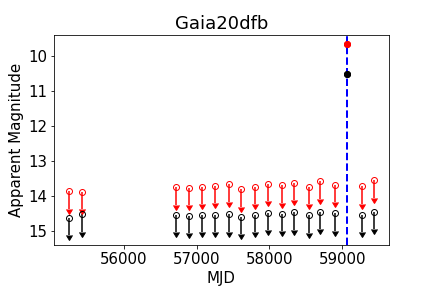}
    \includegraphics[width=0.245\textwidth]{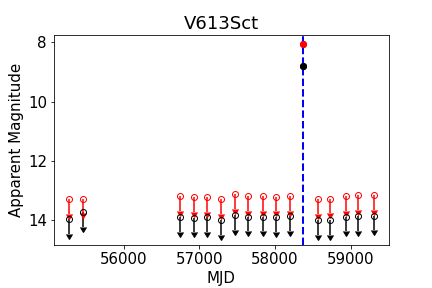}
    \includegraphics[width=0.245\textwidth]{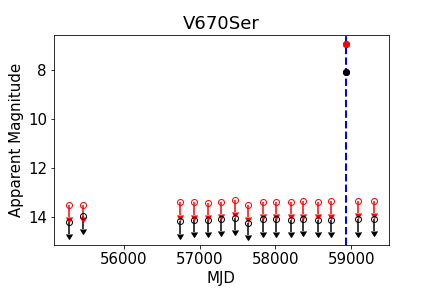}
    \includegraphics[width=0.245\textwidth]{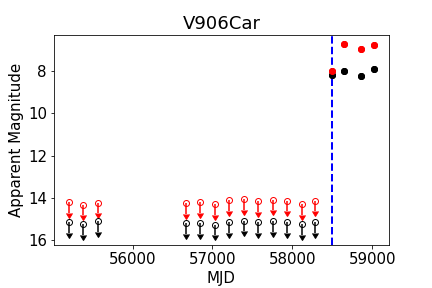}
    \includegraphics[width=0.245\textwidth]{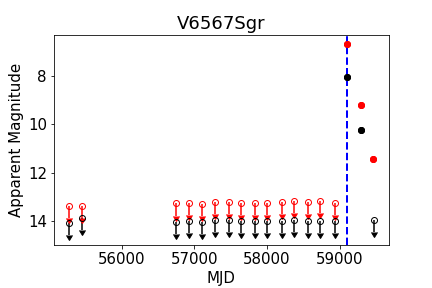}
\caption{WISE forced difference photometry light curves of all known Galactic novae within our search period that did not pass our selection criteria.}
\end{figure}